\documentclass[manuscript]{aastex}
\usepackage{emulateapj5}

\usepackage[dvips]{color}
\usepackage{epsfig}
\usepackage{amsmath}
\usepackage{amssymb}
\usepackage{natbib}
\usepackage{graphicx}

\definecolor{redish}{rgb}{0.7,0.2,0.0}  
\definecolor{bluish}{rgb}{0.2,0.5,0.8}

\shorttitle{Thermonuclear X-ray bursts from 4U 1728--34}
\shortauthors{Bhattacharyya et al.}

\begin{document}

\title{Effects of thermonuclear X-ray bursts on non-burst emissions in the soft state of 4U 1728--34}

\author{Sudip Bhattacharyya\altaffilmark{1}, J. S. Yadav\altaffilmark{1}, Navin Sridhar\altaffilmark{2}, 
	Jai Verdhan Chauhan\altaffilmark{1}, P. C. Agrawal\altaffilmark{3}, H. M. Antia\altaffilmark{1}, 
	Mayukh Pahari\altaffilmark{4}, Ranjeev Misra\altaffilmark{5}, 
	Tilak Katoch\altaffilmark{1}, R. K. Manchanda\altaffilmark{6}, and 
Biswajit Paul\altaffilmark{7}}

\altaffiltext{1}{Department of Astronomy and Astrophysics, Tata Institute of Fundamental Research, Mumbai 400005, India; sudip@tifr.res.in}
\altaffiltext{2}{Indian Institute of Science Education and Research, Bhauri, Bhopal 462066, India}
\altaffiltext{3}{UM-DAE Center of Excellence for Basic Sciences, University of Mumbai, Kalina, Mumbai 400098, India}
\altaffiltext{4}{Royal Society-SERB Newton Fellow, Department of Physics and Astronomy, University of Southampton, Southampton, SO17 3RT, UK}
\altaffiltext{5}{Inter-University Centre for Astronomy and Astrophysics, Pune 411007, India}
\altaffiltext{6}{University of Mumbai, Kalina, Mumbai 400098, India}
\altaffiltext{7}{Department of Astronomy and Astrophysics, Raman Research Institute, Bengaluru 560080, India}

\begin{abstract}
It has recently been shown that the persistent emission of a neutron star low-mass X-ray binary (LMXB)
evolves during a thermonuclear (type-I) X-ray burst. The reason of this evolution, however, 
is not securely known. This uncertainty can introduce significant systematics in the neutron star
radius measurement using burst spectra, particularly if an unknown but significant fraction of the burst emission,
which is reprocessed, contributes to the changes in the persistent emission during the burst.
Here, by analyzing individual burst data of {AstroSat}/LAXPC from the neutron star LMXB 4U 1728--34 in 
the soft state, we show that the burst emission is not significantly reprocessed
by a corona covering the neutron star. Rather,
our analysis suggests that the burst emission enhances the accretion disk emission,
possibly by increasing the accretion rate via disk.
This enhanced disk emission, which is Comptonized by a corona covering the disk, 
can explain an increased persistent emission observed during the burst.
This finding provides an understanding of persistent emission components, and 
their interaction with the thermonuclear burst emission. Furthermore, since burst 
photons are not significantly reprocessed, non-burst and burst emissions can be
reliably separated, which is required to reduce systematic uncertainties 
in the stellar radius measurement.
\end{abstract}


\section{Introduction}\label{Introduction}

Thermonuclear (type-I) X-ray bursts are intermittently observed from many neutron star
low-mass X-ray binaries (LMXBs; \citet{StrohmayerBildsten2006} and references therein).
Such a burst originates from an unstable thermonuclear burning of the accreted matter accummulated
on the neutron star surface \citep{Joss1977, LambLamb1978, StrohmayerBildsten2006}.
For most bursts, the observed X-ray intensity rises in $\approx 0.5-5$ s, decays in $\sim 10-100$ s
as the neutron star surface cools down after the nuclear burning, and
the typical recurrence time is a few hours to days \citep{Galloway2008}. The burst spectrum is
traditionally described with a blackbody model, and the best-fit burst blackbody normalization,
which can be identified as the burst emission area, usually matches well with the expected surface area
of a neutron star \citep{Swank1977, Hoffmanetal1977}. These motivated an effort to measure
the neutron star radius using the burst continuum spectrum, where the normalization of the 
burst blackbody is expected to be proportional to the square of the stellar radius 
\citep[e.g., ][]{Paradijs1978, Goldman1979, Paradijs1979, ParadijsLewin1986}. 
Note that such a radius measurement is extremely important to probe the superdense 
and degenerate core matter of neutron stars, which is a fundamental problem of physics
\citep[e.g., ][]{LattimerPrakash2007, Bhattacharyyaetal2017}. However, a reliable radius 
measurement using this method has so far not been possible due to a number of systematic
uncertainties, which include (1) burst emission from and the visibility of an unknown fraction of
the neutron star surface, (2) plausible deviation of the
burst spectrum from a blackbody, etc. \citep{Bhattacharyya2010, Bhattacharyyaetal2010, Kajavaetal2017a}.
Nevertheless, the use of continuum burst spectrum remains a promising
method to measure the neutron star radius for the following reasons.
(1) Neutron star LMXBs provide
many complementary methods to measure neutron star parameters, the joint application of
which has a potential to significantly reduce the systematics \citep{Bhattacharyya2010}.
(2) It was generally believed that the much more intense burst emission could be reliably 
distinguished from the persistent (i.e., non-burst) emission during the burst.

In the conventional method of distinguishing the burst and the persistent emissions,
the latter is assumed unevolved during the burst, and the observed
preburst spectrum is used as the background while fitting the spectrum during the burst \citep{Galloway2008}.
Then the net spectrum is usually fit with a blackbody model to estimate the burst temperature and
the neutron star radius. But \citet{Worpeletal2013, Worpeletal2015} showed that the persistent flux changes
during bursts. These authors fit the observed spectrum during a burst with 
the scaled (with a free factor $f_a$) frozen best-fit model preburst spectrum plus a blackbody.
This `$f_a$ method', with best-fit $f_a$ values typically between zero and somewhat above 10, 
usually gave a better fit than the above-mentioned conventional method.
This finding raised the following questions.
(1) What is the true nature of the persistent X-ray emitting components, 
and how does a thermonuclear burst interact with them?
(2) Does the shape of the persistent spectrum also evolve during a burst? Since the integration
time for a burst spectrum is typically about a second, the statistics is not usually enough to 
investigate this.
(3) What fraction of X-ray emission during a burst is actually the burst emission, and are the
burst photons reprocessed to contribute to the enhanced persistent emission?
These questions have significant implications for separating the burst emission from the
persistent emission, and hence for a reliable measurement of neutron star radius.

There are also other reports on the persistent emission change during thermonuclear bursts.
For example, \citet{Jietal2013} found a shortage of hard X-ray ($30-50$~keV) photons during bursts
in the hard spectral state of the neutron star LMXB 4U 1636--536.
This was confirmed for other neutron star LMXBs, and was proposed to be due to a corona
cooling by an enhanced number of soft X-ray seed photons from the burst 
\citep{Chenetal2013, Jietal2014a, Jietal2014b, Jietal2015}.
In fact, a decrease in coronal temperature has been detected during the hard state of 
the neutron star LMXB 4U 1728--34,
but only by stacking many burst spectra in the $1-10$ s intervals, in which the burst temperature
significantly evolves \citep{Kajavaetal2017b}. However, such a hard X-ray deficit is not found in 
the soft spectral state of neutron star LMXBs \citep[e.g., ][]{Jietal2013}.

In this paper, for the first time to the best of our knowledge, 
we attempt to understand the reason of persistent emission evolution during 
thermonuclear bursts in the soft state of 4U 1728--34. 
This understanding requires spectral evolution study for individual bursts
(i.e., without stacking them), because both the persistent and the burst spectra evolve
during a burst, and properties of different bursts are usually sufficiently different.
We, for the first time, use the Large Area X-ray Proportional Counter (LAXPC)
instrument aboard the {\it AstroSat} space mission \citep{Yadavetal2016, Antiaetal2017, Agrawaletal2017}
for such a science goal.
Note that LAXPC, which has a large area and does not suffer from pile-up, is ideal to track
burst spectral evolution.
Two proposed models for an enhanced persistent emission during a burst 
in the soft spectral state are based on the following scenarios:
(1) an increased mass accretion rate due to Poynting--Robertson drag on the accretion disk 
because of burst luminosity \citep{Worpeletal2013}, and
(2) a spreading layer, which reprocesses the burst emission and covers a larger extent 
of the neutron star surface during a burst \citep{Kajavaetal2017a}.
The latter model implies a burst emission region almost entirely covered by a
Comptonizing spreading layer, consequently an unknown fraction of burst photons
being Comptonized, and hence a larger uncertainty in the neutron star radius measurement.
Our systematic analysis suggests that the former model is more viable.

\section{Observations and spectral state}\label{Observations}

The persistent neutron star LMXB 4U 1728--34 was observed with {\it AstroSat}/LAXPC in 
the performance verification phase in March 2016. The observations (obsId: 
T01\_041T01\_9000000362) spanned over 22 satellite orbits from 07 March 2016 16:53:11 to
09 March 2016 09:40:00. In the total 61.05 ks of useful LAXPC exposure, five
thermonuclear bursts were detected. The times of these bursts, using 
10\% of the peak count rate for both start and end times, are (1) 08 March 2016 02:58:39.2--02:58:50.2, (2) 08 March 2016 
13:52:10.0--13:52:23.4, (3) 08 March 2016 16:43:26.5--16:43:37.8, (4) 08 March 2016
23:25:24.0--23:25:36.1 and (5) 09 March 2016 03:13:56.1--03:14:08.8.
Note that high frequency timing features have already been reported from these LAXPC data 
\citep{VerdhanChauhanetal2017}. In this paper, however, 
we focus on the spectral evolution during the five thermonuclear bursts. For this, 
we generate the total spectrum, the background spectrum and the response file 
using the laxpc software\footnote{http://www.tifr.res.in/$\sim$astrosat\_laxpc/LaxpcSoft.html}
\citep{Antiaetal2017}.

\begin{figure}[h]
\centering
\includegraphics*[width=14.0cm]{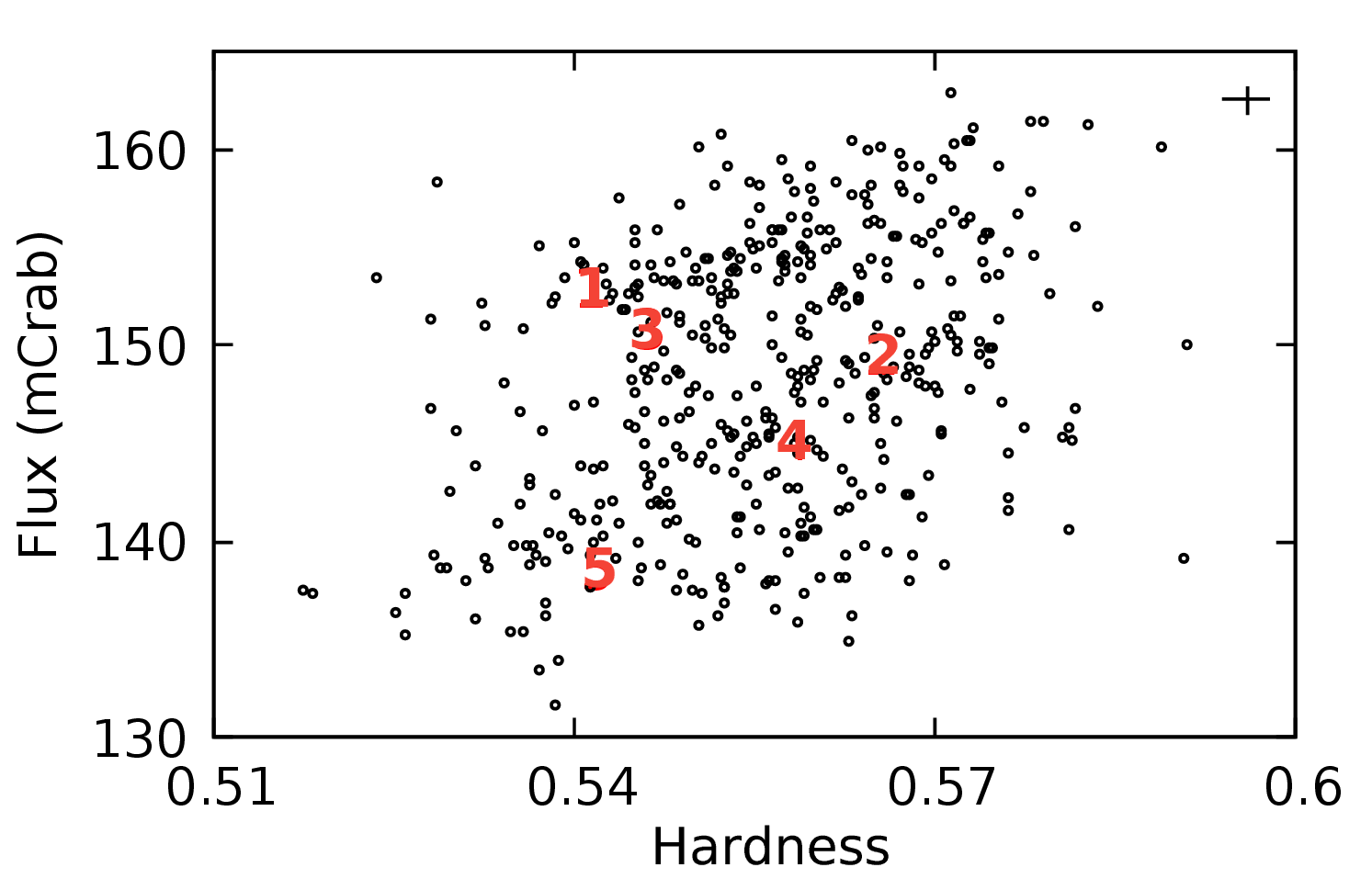}
\caption{Hardness-intensity diagram of the neutron star LMXB 4U 1728--34 using {\it AstroSat}/LAXPC
data of March, 2016. The definitions of the hardness and intensity are given in
Section~\ref{Observations}. Each point is for 128 s integration time, typical error bars for both
hardness and intensity are shown, and red numbers represent persistent emissons just before thermonuclear X-ray bursts.
This figure shows that 4U 1728--34 was in a high intensity soft spectral state.
\label{fig1}}
\end{figure}

In order to interpret the results, it is useful to know the spectral state of the source 
during these observations. We, therefore, make the hardness-intensity diagram (HID) of 
4U 1728--34 using the data from one LAXPC detector (LAXPC10) in the following way. 
The hardness is considered to be the ratio of background-subtracted LAXPC10
count rate in $8-25$~keV to that in $3-8$~keV. We also estimate this ratio for the closest LAXPC10
observation of Crab, for which the LAXPC10 gain was the same as for 4U 1728--34 observations, 
and normalize the 4U 1728--34 hardness with the Crab hardness (Fig.~\ref{fig1}). 
The intensity of the HID is defined as the background-subtracted $3-25$~keV 
LAXPC10 count rate 
(typically $\sim 370$) 
of 4U 1728--34 normalized by that 
($\sim 2491$) 
of Crab. 
Note that expressing intensity and hardness in the Crab unit is useful, to meaningfully compare HIDs 
from the data of different instruments, and even from the data of different gain epochs 
of the same instrument \citep{Bhattacharyya2007}.
Fig.~\ref{fig1} is for the 61.05 ks of useful LAXPC exposure, excluding the durations of
five bursts, and each point is for a 128 s integration time. Each of the five bursts is represented 
in the HID by the 128 s duration point just before the burst.
Comparing our Fig.~\ref{fig1} with Fig.~3 of \citet{Kajavaetal2017b}, we can conclude that 
4U 1728--34 was in a high intensity soft spectral state during the {\it AstroSat}/LAXPC observations.

\section{Spectral analysis and results}\label{Spectral}

For spectral analysis, we use LAXPC10 and LAXPC20, and exclude the third LAXPC unit (LAXPC30)
due to its gain instability caused by a gas leakage.
The dead time effect is taken care of by the correction of the count rate by the laxpc software.
A 2\% systematic error is applied for the fitting of spectra.
Since LAXPC cannot characterize the source below 3 keV, we use a fixed hydrogen column density 
of $2.3\times10^{22}$ cm$^{-2}$ \citep{Gallowayetal2010} in the multiplicative 
XSPEC software\footnote{https://heasarc.gsfc.nasa.gov/docs/xanadu/xspec/} model component {\tt tbabs}.
The 90\% errors of parameters are estimated using the MCMC technique with 
the Goodman-Weare algorithm, a chain length of 100000 steps (30000 in the initial burn-in phase)
and 1000 walkers \citep[e.g., ][]{Ingrametal2016}.

The persistent or non-burst spectral components of a neutron star LMXB can be the following
\citep{MukherjeeBhattacharyya2011}:
(1) a geometrically thin accretion disk emitting a multi-color blackbody or disk blackbody spectrum,
(2) a boundary layer on the neutron star emitting a (single temperature) blackbody spectrum, and
(3) one or more coronae (hot electron gas), which may partially or entirely cover the disk and/or 
the boundary layer, reprocessing (Comptonizing) the radiations from these components.
Here, however, we aim to fit the preburst spectrum of 200 s exposure 
for each of five bursts with the simplest model which gives 
a reasonable value of reduced $\chi^2$ ($\chi^2$/dof; `dof': degrees of freedom).
This is because our main goal is to study how the parameter values of a preburst spectrum evolve during the burst,
and since there are relatively small number of counts within a $\sim 1$~s time bin during a burst, 
such a study could be feasible only if the number of preburst spectral parameters is small.

The simplest physical model, which describes the preburst spectra well, is a Comptonization model, and we
choose it to be {\tt nthcomp} \citep{Zdziarskietal1996, Zyckietal1999} in XSPEC.
The {\tt nthcomp} offers two options: the seed photons can be supplied by a single temperature blackbody 
(e.g., a boundary layer) or by a disk blackbody (e.g., an accretion disk). Hence,
the sole {\tt nthcomp} model component implies a Comptonizing layer (corona) entirely
covering either a boundary layer or a disk, none of which is directly observed.
Therefore, the {\tt nthcomp} model is very convenient in the sense that it effectively represents
two spectral components with small number of parameters. In Table~\ref{preburst}, we show the results
of spectral fits with both options for the preburst spectrum of burst-1. The {\tt nthcomp} electron temperature is relatively 
low in both cases, which is expected as the source was in a soft state. Both the models are almost 
equally preferred based on the $\chi^2$/dof values, and hence from these fits we cannot 
conclude if the Comptonizing layer primarily covers the disk or the neutron star. However, 
we find that the seed photon temperature for a 
single temperature blackbody is significantly lower than that for a disk blackbody. These results are
qualitatively same for four other preburst spectra.

Then we probe the spectral evolutions during bursts. For this, we divide each burst duration into several time bins, 
so that each bin ($\sim 1-5$~s) has at least 5000 background-subtracted counts, and perform spectral
fitting for each time bin. First, we follow the conventional method of fitting, in which the preburst spectrum
is used as the background spectrum (see Section~\ref{Introduction}). The results for burst-1 are shown in
Table~\ref{conventional}. Then we use the so-called `$f_a$ method' (see Section~\ref{Introduction}) 
for each of the two preburst models for burst-1, and Table~\ref{conventional} shows the results.
As also reported earlier \citep[e.g., ][]{Worpeletal2013}, the `$f_a$ method' provides a better fit
than the conventional method. 
We notice that the persistent emission during burst-1, which first increases and then
decreases, is always higher than the preburst emission. Moreover, the burst blackbody normalization,
which is used to estimate the neutron star radius (Section~\ref{Introduction}), is usually
lower for the `$f_a$ method' than for the conventional method. This shows that the conventional method
can introduce systematic error in the estimated stellar radius.
However, the $f_a$ value and burst properties are very similar for the two preburst models, and hence,
even from the Table~\ref{conventional}, we cannot conclude if the Comptonizing layer mainly
covers the disk or the neutron star. We note that very similar results are obtained also from the 
four other bursts.

However, the `$f_a$ method' mentioned above, which assumes that the persistent spectral shape remains
unchanged during a burst, may not be self-consistent in some cases. Suppose, the Comptonizing layer partially
covers the neutron star during a burst. In such a case, a fraction of the burst photons can directly 
be detected by the observer,
and the rest can be reprocessed by the Comptonizing layer giving rise to the {\tt nthcomp}
spectral component. Hence, the burst temperature, which
can be obtained from the directly observed burst blackbody, should be same as the seed photon 
temperature of {\tt nthcomp}. Consequently, this seed photon temperature would be different from the
preburst value, and the persistent spectral shape would change during a burst.
Therefore, next we follow a modified `$f_a$ method' starting from the results 
given in the middle section 
(the case of blackbody seed photons for {\tt nthcomp}) of Table~\ref{conventional}. 
We freeze $f_a$, as well as the burst blackbody temperature and normalization, make the 
frozen {\tt nthcomp} seed photon temperature same as the burst blackbody temperature,
and thaw the {\tt nthcomp} photon index and electron temperature. 
However, the electron temperature should not be less than the seed photon temperature, and 
should not be more than the preburst electron temperature, as the burst photons may
cool down the corona \citep{Kajavaetal2017b}.
Therefore, while fitting, we constrain the electron temperature in the range of 
burst blackbody temperature and preburst electron temperature.
These gave unaccepted fits for all time bins  of all the five bursts 
($\chi^2$/dof $\sim 3-800$, and typically $\sim 100$). Since, here we assume that an
unknown fraction of burst photons are reprocessed by the Comptonizing layer,
it is reasonable to thaw both $f_a$ and the burst blackbody normalization.
The results with such thawing for burst-1 are given in the first section of Table~\ref{final}.
The $\chi^2$/dof values, which are mostly above $2$, show that fits are 
generally unacceptable. In fact, it was not possible to estimate the errors
of the best-fit parameters in some cases.
The best-fit {\tt nthcomp} electron temperature also pegs at the
{\tt nthcomp} seed photon temperature for most cases, which implies essentially no Comptonization
of burst photons. This is consistent with the fact that $f_a$, and hence the flux from the
Comptonizing layer, are very low (Table~\ref{final}). 
Similar results, i.e., generally high $\chi^2$/dof values and very low flux from the Comptonizing layer,
are also found for the other four bursts. These show that burst
photons are not significantly reprocessed by a Comptonizing layer, and a Comptonizing layer
covering the neutron star cannot explain the observed spectra during bursts. 

Now, we probe the other scenario, i.e., the Comptonizing layer covering the disk
and reprocessing its emission. Since our spectral model does not explicitly include a disk
component, we cannot tie a disk blackbody temperature with the {\tt nthcomp} seed photon temperature.
So we start from the results given in the last section of 
Table~\ref{conventional}, and thaw the {\tt nthcomp} photon index, seed photon temperature
and electron temperature. However, as explained earlier, we constrain the electron temperature
at or below its preburst value. 
We also freeze the best-fit burst blackbody temperature 
of Table~\ref{conventional}.
But we keep $f_a$, which determines the overall normalization of the persistent {\tt nthcomp}
component, as a free parameter.
The results of these fits for burst-1
are given in the last section of Table~\ref{final}. The fits are acceptable, and the $\chi^2$/dof values
are clearly much smaller than those for the reprocessing of burst emission case (given in the
first section of Table~\ref{final}). 
These results are generally similar for other four bursts, and show that the scenario of
the Comptonizing layer covering the disk and reprocessing the disk emission is favored. Therefore,
it is likely that the enhanced persistent emission during a burst is due to an increased accretion
rate via disk (see Section~\ref{Introduction}).

Note that an increased accretion rate may imply a higher disk temperature during the burst, 
and generally a persistent spectral shape evolved from the preburst spectral shape.
However, due to large errors of best-fit spectral parameter values, we can neither establish
a persistent spectral shape evolution, nor rule out this possibility (Table~\ref{preburst}
and Table~\ref{final}).
Besides, a part of the enhanced persistent emission could be due to the reflection
of burst photons from the disk. While the relatively small number of observed counts during 
a burst time bin does not allow us to test this rigorously, we did attempt to include a broad Gaussian emission
component, representative of the broad Fe K$\alpha$ emission line, as a proxy to the 
reflection spectral component. This is because this line can be the most prominent feature in the
reflection spectrum \citep{Fabianetal2000, BhattacharyyaStrohmayer2007}. Since this Gaussian component is not significant 
(e.g., $< 2\sigma$ for burst-1 time bins), we conclude that the reflection of burst photons from 
the disk may not be substantial in this case.

\section{Discussion}\label{Discussion}

In this paper, we have analyzed individual thermonuclear X-ray bursts from 4U 1728--34 in the 
soft state observed with a new X-ray instrument {\it AstroSat}/LAXPC.
We show that the burst emission is not significantly reprocessed by a Comptonizing layer 
(e.g., a spreading boundary layer) covering the neutron star.
Rather, the burst emission possibly increases the disk emission by enhancing the accretion rate via disk,
perhaps due to Poynting--Robertson drag on the accretion disk. 
This increased disk emission, Comptonized by a corona covering the disk, provides
an enhanced persistent emission during a burst. This conclusion has the following implications.
(1) It provides an understanding of persistent emission components, and suggests that the corona
covers the accretion disk and reprocesses disk emision.
(2) It provides an understanding of effects of the burst emission on persistent emission in the soft state.
(3) Since the burst emission is not significantly reprocessed by the corona, this emission
could be reliably distinguished from the persistent or non-burst emission, which is required to
reduce systematics in neutron star radius measurements using burst spectra.
This is important because stellar radius measurement by this method is a goal of future X-ray instruments.
Finally, we note that the burst ignition latitude could be
inferred from the shape of the burst rising light curve \citep{MaurerWatts2008, ChakrabortyBhattacharyya2014}.
We find that this shape depends on which method of spectral fitting is used.

\acknowledgments

We acknowledge the LAXPC Payload Operation Center (TIFR, Mumbai) and 
Indian Space Research Organization for their support.

\begin{table*}
\centering
        \caption{Best-fit parameters (for XSPEC model {\tt tbabs*nthcomp}) of the 200 s
        preburst {\it AstroSat}/LAXPC spectrum of burst-1 from 4U 1728--34 (Section~\ref{Spectral}).}
\begin{tabular}{lcc}
\hline
        Spectral & Blackbody & Disk blackbody \\
        parameters & seed photons\footnotemark[1] & seed photons\footnotemark[2] \\
\hline
        $n_{\rm H}$\footnotemark[3] & 2.3 (frz\footnotemark[4]) & 2.3 (frz)\\
        $\Gamma$\footnotemark[5] & $1.93_{-0.05}^{+0.07}$ & $2.15_{-0.14}^{+0.23}$\\
        $kT_{\rm e}$\footnotemark[6] & $3.17_{-1.06}^{+0.83}$ & $3.45_{-1.06}^{+0.51}$\\
        $kT_{\rm seed}$\footnotemark[7] & $0.82_{-0.06}^{+0.06}$ & $1.63_{-0.19}^{+0.22}$\\
        Norm\footnotemark[8] & $0.12_{-0.01}^{+0.02}$ & $0.27_{-0.01}^{+0.01}$\\
        Obs. flux\footnotemark[9] & 2.83 & 2.83\\
        Unabs. flux\footnotemark[10] & 3.88 & 4.38\\
        Reduced $\chi^2$ (dof) & 1.12 (122) & 1.08 (122)\\
\hline
\end{tabular}
\begin{flushleft}
$^1$ Seed photons for the thermal Comptonization component ({\tt nthcomp}) are provided by a blackbody.\\
$^2$ Seed photons for {\tt nthcomp} are provided by a disk blackbody.\\
$^3$ Hydrogen column density (in $10^{22}$ cm$^{-2}$) for {\tt tbabs}.\\
$^4$ ``frz" indicates that the parameter value is frozen.\\
$^5$ Asymptotic power-law photon index for {\tt nthcomp}.\\
$^6$ Electron temperature (in keV) for {\tt nthcomp}.\\
$^7$ Seed photon temperature (in keV) for {\tt nthcomp}.\\
$^8$ Normalization for {\tt nthcomp}.\\
$^9$ Observed flux (in $10^{-9}$ erg cm$^{-2}$ s$^{-1}$) in $3-25$~keV.\\
$^{10}$ Unabsorbed bolometric flux (in $0.01-100$~keV using the best-fit model parameters; in $10^{-9}$ erg cm$^{-2}$ s$^{-1}$).
\end{flushleft}
\label{preburst}
\end{table*}

\begin{table*}
\centering
\caption{Best-fit parameters of {\it AstroSat}/LAXPC spectra of the five time bins of burst-1
from 4U 1728--34 with unevolved shape of the persistent spectrum during the burst
(Section~\ref{Spectral}).}
\begin{tabular}{lccccc}
\hline
Spectral & Time bin 1 & Time bin 2 & Time bin 3 & Time bin 4 & Time bin 5\\
parameters & & & & & \\
\hline
\multicolumn{6}{c}{Conventional method of burst spectral fitting (with XSPEC model {\tt tbabs*bbodyrad})}\\
\hline\\
$kT_{\rm BB}$\footnotemark[1] & $2.47_{-0.04}^{+0.04}$ & $2.51_{-0.03}^{+0.03}$ & $2.20_{-0.04}^{-0.04}$ & $1.86_{-0.04}^{+0.04}$ & $1.55_{-0.05}^{+0.05}$ \\
Norm$_{\rm BB}$\footnotemark[2] & $44.58_{-2.97}^{+3.18}$ & $62.64_{-3.36}^{+3.55}$ & $85.58_{-5.51}^{+5.90}$ & $85.17_{-7.20}^{+7.93}$ & $86.37_{-10.57}^{+12.16}$ \\
Unabs. flux & $17.74$ & $26.96$ & $21.53$ & $11.05$ & $5.32$ \\
Reduced $\chi^2$ (dof) & 1.42 (124) & 1.36 (124) & 1.53 (124) & 2.41 (120) & 1.79 (124) \\
\hline
\multicolumn{6}{c}{$f_a$ method of burst spectral fitting (for blackbody seed photons for {\tt nthcomp})}\\
\hline\\
$f_a$ & $2.30_{-0.42}^{+0.40}$ & $2.41_{-0.50}^{+0.49}$ & $3.27_{-0.60}^{+0.58}$ & $3.80_{-0.40}^{+0.38}$ & $1.85_{-0.11}^{+0.10}$ \\
$kT_{\rm BB}$\footnotemark[1] & $2.64_{-0.08}^{+0.09}$ & $2.63_{-0.06}^{+0.06}$ & $2.33_{-0.06}^{+0.07}$ & $1.69_{-0.25}^{+0.17}$ & $1.30_{-0.11}^{+0.10}$ \\
Norm$_{\rm BB}$\footnotemark[2] & $26.53_{-5.07}^{+5.80}$ & $43.91_{-6.29}^{+6.96}$ & $45.59_{-9.66}^{+10.73}$ & $24.78_{-12.91}^{+14.16}$ & $94.77_{-21.56}^{+32.51}$ \\
Unabs. flux ({\tt $f_a$*nthcomp})\footnotemark[3] & $8.92$ & $9.35$ & $12.69$ & $14.74$ & $7.18$ \\
Unabs. flux ({\tt bbodyrad})\footnotemark[4] & $13.87$ & $22.61$ & $14.46$ & $2.18$ & $2.91$ \\
Obs. flux & $18.80$ & $27.10$ & $21.87$ & $12.34$ & $7.13$ \\
Reduced $\chi^2$ (dof) & 1.24 (123) & 1.24 (123) & 1.25 (123) & 1.45 (123) & 1.46 (123) \\
\hline
\multicolumn{6}{c}{$f_a$ method of burst spectral fitting (for disk blackbody seed photons for {\tt nthcomp})}\\
\hline\\
$f_a$ & $2.27_{-0.41}^{+0.39}$ & $2.36_{-0.49}^{+0.48}$ & $3.22_{-0.59}^{+0.57}$ & $3.70_{-0.39}^{+0.37}$ & $1.80_{-0.18}^{+0.17}$ \\
$kT_{\rm BB}$\footnotemark[1] & $2.64_{-0.08}^{+0.09}$ & $2.63_{-0.06}^{+0.06}$ & $2.33_{-0.06}^{+0.07}$ & $1.68_{-0.24}^{+0.17}$ & $1.30_{-0.11}^{+0.10}$ \\
Norm$_{\rm BB}$\footnotemark[2] & $26.47_{-5.06}^{+5.79}$ & $43.95_{-6.28}^{+6.95}$ & $45.46_{-9.65}^{+10.73}$ & $26.51_{-12.98}^{+14.36}$ & $96.68_{-21.79}^{+32.81}$ \\
Unabs. flux ({\tt $f_a$*nthcomp})\footnotemark[3] & $9.94$ & $10.34$ & $14.10$ & $16.21$ & $7.88$ \\
Unabs. flux ({\tt bbodyrad})\footnotemark[4] & $13.84$ & $22.63$ & $14.42$ & $2.24$ & $2.93$ \\
Obs. flux & $18.81$ & $27.11$ & $21.88$ & $12.35$ & $7.13$ \\
Reduced $\chi^2$ (dof) & 1.23 (123) & 1.24 (123) & 1.25 (123) & 1.47 (123) & 1.49 (123) \\
\hline
\end{tabular}
\begin{flushleft}
$^1$ Thermonuclear burst blackbody temperature (in keV).\\
$^2$ Thermonuclear burst blackbody normalization.\\
$^{3}$ Unabsorbed bolometric flux of the evolved persistent component {\tt $f_a$*nthcomp} (in $0.01-100$~keV using the best-fit model parameters; in $10^{-9}$ erg cm$^{-2}$ s$^{-1}$).\\
$^{4}$ Unabsorbed bolometric flux of the burst component {\tt bbodyrad} (in $0.01-100$~keV using the best-fit model parameters; in $10^{-9}$ erg cm$^{-2}$ s$^{-1}$).\\
\end{flushleft}
\label{conventional}
\end{table*}

\begin{table*}
\centering
\caption{Best-fit parameters of {\it AstroSat}/LAXPC spectra of the five time bins of burst-1
from 4U 1728--34 with evolved shape of the persistent spectrum during the burst
(Section~\ref{Spectral}).}
\begin{tabular}{lccccc}
\hline
Spectral & Time bin 1 & Time bin 2 & Time bin 3 & Time bin 4 & Time bin 5\\
parameters & & & & & \\
\hline
\multicolumn{6}{c}{Burst blackbody provides seed photons for the evolved persistent component {\tt nthcomp}}\\
\hline\\
$f_a$ & $0.01$ & $0.002$ & $0.002$ & $0.02_{-0.0}^{+0.02}$ & $0.03_{-0.01}^{+0.00}$ \\
$\Gamma$ ({\tt nthcomp}) & 10 & 10 & 10 & $1.00_{-0.0}^{+0.410}$ & $1.00_{-0.0}^{+0.32}$ \\
$kT_{\rm e}$ ({\tt nthcomp}) & 2.64 & 2.63 & 2.33 & $3.17_{-0.10}^{+0.00}$ & $2.64_{-0.12}^{+0.15}$  \\
$kT_{\rm seed}$ ({\tt nthcomp}) &  &  &  &  &  \\
$= kT_{\rm BB}$ ({\tt bbodyrad}) &  &  &  &  &  \\
(frz) & 2.64 & 2.63 & 2.33 & 1.69 & 1.30 \\
Norm$_{\rm BB}$ ({\tt bbodyrad}) & 40.60 & 59.40 & 79.61 & $134.35_{-5.57}^{+3.18}$ & $208.79_{-26.52}^{+10.85}$ \\
Unabs. flux ({\tt $f_a$*nthcomp}) & 0.28 & 0.06 & 0.04 & 3.26 & 2.49 \\
Unabs. flux ({\tt bbodyrad}) & 21.23 & 30.59 & 25.25 & 11.80 & 6.42 \\
Obs. flux & 19.30 & 27.47 & 21.88 & 12.08 & 7.06 \\
Reduced $\chi^2$ (dof) & 2.82 (122) & 2.41 (122) & 2.41 (122) & 2.19 (122) & 1.51 (122) \\
\hline
\multicolumn{6}{c}{The disk (which is not directly seen) provides seed photons for the evolved persistent component {\tt nthcomp}}\\
\hline\\
$f_a$ & $2.20_{-0.28}^{+0.26}$ & $2.27_{-0.31}^{+0.30}$ & $3.16_{-0.39}^{+0.46}$ & $4.06_{-0.76}^{+0.65}$ & $3.36_{-1.04}^{+1.29}$ \\
$\Gamma$ ({\tt nthcomp}) & $2.30_{-0.36}^{+0.42}$ & $2.21_{-0.35}^{+0.56}$ & $2.32_{-0.33}^{+0.45}$ & $2.04_{-0.25}^{+0.25}$ & $1.86_{-0.18}^{+0.24}$ \\
$kT_{\rm e}$ ({\tt nthcomp}) & $3.44_{-0.48}^{+0.01}$ & $3.43_{-0.36}^{+0.02}$ & $3.44_{-0.54}^{+0.01}$ & $3.07_{-0.24}^{+0.31}$ & $2.67_{-0.17}^{+0.29}$ \\
$kT_{\rm seed}$ ({\tt nthcomp}) & $2.64_{-0.96}^{+0.68}$ & $2.49_{-0.85}^{+0.73}$ & $2.18_{-0.62}^{+0.71}$ & $1.59_{-0.64}^{+0.34}$ & $0.95_{-0.62}^{+0.47}$ \\
$kT_{\rm BB}$ ({\tt bbodyrad}) (frz) & 2.64 & 2.63 & 2.33 & 1.68 & 1.30 \\
Norm$_{\rm BB}$ ({\tt bbodyrad}) & $18.18_{-6.64}^{+8.76}$ & $36.26_{-6.13}^{+7.34}$ & $37.08_{-14.59}^{+10.35}$ & $8.64_{-7.08}^{+37.76}$ & $36.47_{-29.93}^{+61.89}$ \\
Unabs. flux ({\tt $f_a$*nthcomp}) & 14.23 & 14.38 & 16.66 & 18.06 & 11.02 \\
Unabs. flux ({\tt bbodyrad}) & 9.5 & 18.67 & 11.76 & 0.74 & 1.12 \\
Obs. flux & 18.85 & 27.14 & 21.91 & 12.34 & 7.13 \\
Reduced $\chi^2$ (dof) & 1.22 (121) & 1.23 (121) & 1.26 (121) & 1.49 (121) & 1.39 (121) \\
\hline
\end{tabular}
\begin{flushleft}
\end{flushleft}
\label{final}
\end{table*}

\clearpage

\end{document}